\def\be{\begin{equation}}
\def\ee{\end{equation}}
\def\ba{\begin{array}}
\def\ea{\end{array}}

\documentclass[aps,pra,preprint,showpacs,amsmath,amssymb,amsfonts,preprintnumbers]{revtex4}
\usepackage{epsfig,graphicx, color}
\usepackage{amsmath}

\newtheorem{theorem}{Theorem}

\begin{document}

\baselineskip=22pt \setcounter{page}{1} \centerline{\Large\bf
On Uncertainty Relations in the Product Form} \vspace{6ex}
\begin{center}
 Xiaofen Huang$^{1}$, Tinggui Zhang$^{1\ast}$,  Naihuan Jing$^{2,3}$
\bigskip

\begin{minipage}{6.5in}
$^1$ School of Mathematics and Statistics, Hainan Normal University, Haikou 571158, China\\
$^2$ School of Mathematics, South China University of Technology, Guangzhou 510640, China\\
$^3$ Department of Mathematics, North Carolina State University,
Raleigh, NC27695, USA\\
$\ast$Corresponding author: tinggui333@163.com
\end{minipage}
\end{center}
\vskip 1 true cm
\begin{center}
\begin{minipage}{5in}
\vspace{3ex} \centerline{\large Abstract} \vspace{4ex}

We study the uncertainty relation in the product form of variances
and obtain some new uncertainty relations with weight, which are shown to be tighter than those 
derived from the Cauchy-Schwarz inequality.

\end{minipage}
\end{center}



\section{Introduction}
The uncertainty relations have played a fundamental role in the development of quantum theory not only in the foundation and also
in recent investigations of quantum information and quantum communication, in particular in the areas such as
entanglement detection \cite{entan1, entan2}, security analysis of quantum key distribution in quantum cryptography \cite{secu}, quantum metrology and quantum speed limit \cite{qsl1, qsl2, qsl3}. Usually the uncertainty relations are expressed in terms of the product of variances of the measurement results of two incompatible observables. Other forms of the uncertainty relations
include  
entropic uncertainty principle \cite{ent1, ent2, ent3}, the majorization technique \cite{maj1, maj2, maj3} and the recent weighted uncertainty relation \cite{XJLF}.

In 1927 Heisenberg \cite{heisen} analyzed the observation of an individual electron with photons and obtained the famous uncertainty principle
\begin{equation}\label{heis}
    (\Delta P)^2(\Delta Q)^2\geq(\frac{\hbar}{2})^2,
\end{equation}
where $(\Delta P)^2$ and $(\Delta Q)^2$ are the variances of the position $P$ and momentum $Q$ respectively. The variance or standard
deviation of the observable $A$ with respect to the state $\rho$ is defined by $(\Delta A)^2=\langle A^2\rangle-\langle A\rangle^2$, where $\langle A\rangle=\mathrm{tr} \rho A$ is the mean value of the observable $A$.

 The inequality (\ref{heis}) shows that the uncertainty in the position and momentum of a quantum particle are inversely proportional to each other: a particle's position and momentum cannot be known simultaneously. Thus the accuracy of quantum measurement is limited by the uncertainty principle. This principle uncovers a fundamental and peculiar feature in the atomic world, and is
 considered as one of the cornerstones of quantum mechanics.

Robertson \cite{robert} formulated the uncertainty relation for arbitrary pair of non-commuting observables $A$ and $B$ (with bounded spectra):
\begin{equation}\label{robert}
    (\Delta A)^2(\Delta B)^2\geq \frac{1}{4}|\langle[A, B]\rangle|^2.
\end{equation}
where $\langle[A, B]\rangle=\mathrm{tr}\rho [A, B]$, the expectation value of the
commutator $[A, B]=AB-BA$. 
Robertson's uncertainty relation is further generalized by Schr\"{o}dinger \cite{sch}:
\begin{equation}\label{sch}
     (\Delta A)^2(\Delta B)^2\geq\frac{1}{4}|\langle[A, B]\rangle|^2+\frac{1}{4}|\langle\{A, B\}\rangle-\langle A\rangle \langle B\rangle|^2.
\end{equation}
This relation is evidently stronger than Heisenberg's uncertainty relation, and it also shows that
the commutator reveals incompatibility while the anticommutator encodes correlation between observables $A$ and $B$.

The goal of this note is to give a family of generalized Schr\"odinger uncertainty relations using
a stronger Cauchy-Schwarz inequality.

%

\section{Generalized Uncertainty Relations}
We start with a quantum system in the quantum state $\rho=|\Psi\rangle\langle\Psi|$
on the Hilbert space with the inner product $\langle \ |\ \rangle$ 
and consider observables $A$ and $B$. Define
 the operator $\bar{A}=A-\langle A\rangle I$ associated with a given operator $A$.
Let $\{|\varphi_i\rangle\}$ be an orthonormal basis of the Hilbert space and write
 $\bar{A}|\Psi\rangle=\sum_i \alpha_i |\varphi_i\rangle$, $\bar{B}|\Psi\rangle=\sum_i \beta_i |\varphi_i\rangle$
 so that
 $\langle\bar{A}\bar{B}\rangle=\langle\alpha|\beta\rangle$, where $\langle\alpha|\beta\rangle$ is
 the usual inner product (linear in the second argument) for the vectors
$\alpha=(\alpha_1, \alpha_2, \ldots, \alpha_n)$ and $\beta=(\beta_1, \beta_2, \ldots, \beta_n)$. We will not distinguish
the two inner products as long as it is clear from the context.

The variance of observable $A$ can be expressed as $(\Delta A)^2=\langle A^2\rangle-\langle A\rangle^2=\langle \bar{A}^2\rangle= \langle \alpha|\alpha\rangle$, thus
$$ (\Delta A)^2(\Delta B)^2= \langle \alpha|\alpha\rangle \langle \beta|\beta\rangle
=\sum_{i=1}^{n} |\alpha_i|^2\sum_{i=1}^{n} |\beta_i|^2.
$$

\begin{theorem} For observables $A$ and $B$, we have the generalized uncertainty relation in the variance-based product form given by
\begin{equation}\label{res1}
(\Delta A)^2(\Delta B)^2\geq \sum_{i=1}^{n}|\alpha_i|^{1+\lambda}|\beta_i|^{1-\lambda}\sum_{i=1}^{n}|\alpha_i|^{1-\lambda}|\beta_i|^{1+\lambda},
\end{equation}
where $\lambda$ is any real number $\in [0, 1]$, $\alpha_i=\langle\varphi_i|\bar{A}|\Psi\rangle$ and $\beta_i=\langle\varphi_i|\bar{B}|\Psi\rangle$.
\end{theorem}
Proof: For any real number $\lambda\in [0, 1]$ one has the following Callebaut inequality \cite{calle}:
\begin{equation}\label{1}
(\sum_{i=1}^{n} a_i b_i)^2\leq \sum_{i=1}^{n}a_i^{1+\lambda}b_i^{1-\lambda}\sum_{i=1}^{n}a_i^{1-\lambda}b_i^{1+\lambda}\leq  \sum_{i=1}^{n}a_i^2 \sum_{i=1}^{n}b_i^2.
\end{equation}
where $\{a_i\}_{i=1}^n$, $\{b_i\}_{i=1}^n$ are two sequences of positive real numbers.
Then
$$ (\Delta A)^2(\Delta B)^2=\sum_{i=1}^{n} |\alpha_i|^2\sum_i^n |\beta_i|^2\geq \sum_{i=1}^{n}|\alpha_i|^{1+\lambda}|\beta_i|^{1-\lambda}\sum_{i=1}^{n}|\alpha_i|^{1-\lambda}|\beta_i|^{1+\lambda},
$$
so we get the inequality in the Theorem.

We remark that our generalized uncertainty relation is stronger than Schr\"odinger's uncertainty relation,
as the generalized Cauchy inequality shows that $|\langle \alpha|\beta\rangle|$ is smaller than
the right-hand side of our uncertainty relation.

The uncertainty relation \eqref{res1} can be tightened by optimizing over the sets of complete orthonormal bases.
Then we can improve the uncertainty relation by the Callebaut inequality as in the proof of Theorem 1.

\begin{theorem}
For observables $A$ and $B$, one has the following uncertainty relation
\begin{equation}\label{th11} 
(\Delta A)^2(\Delta B)^2\geq \max\limits_{\{|\varphi_i\rangle\}} 
\sum_{i=1}^{n}|\alpha_i|^{1+\lambda}|\beta_i|^{1-\lambda}\sum_{i=1}^{n}|\alpha_i|^{1-\lambda}|\beta_i|^{1+\lambda}=\mathcal{L}_1.
\end{equation}
where $\lambda$ is any fixed number $\in [0, 1]$,
 $\alpha_i=\langle\varphi_i|\bar{A}|\Psi\rangle$ and $\beta_i=\langle\varphi_i|\bar{B}|\Psi\rangle$.
\end{theorem}


\begin{theorem}
For observables $A$ and $B$, we can get the uncertainty relation in the following
\begin{equation}\label{th2}
(\Delta A)^2(\Delta B)^2\geq \max\limits_{\{|\varphi_i\rangle\}}\sum_{i=1}^{n}(|\alpha_i|^{2}+|\beta_i|^{2})
\sum_{i=1}^{n}\frac{|\alpha_i|^{2}|\beta_i|^{2}}{|\alpha_i|^{2}+|\beta_i|^{2}}
:=\mathcal{L}_2,
\end{equation}
where  $\alpha_i=\langle\varphi_i|\bar{A}|\Psi\rangle$ and $\beta_i=\langle\varphi_i|\bar{B}|\Psi\rangle$.
\end{theorem}

{\bf Proof}: Consider another generalized Cauchy-Schwarz inequality, the Milne inequality \cite{Mil}:
\begin{equation}\label{3}
(\sum_{i=1}^{n} a_i b_i)^2\leq \sum_{i=1}^{n}(a_i^{2}+b_i^{2})\sum_{i=1}^{n}\frac{a_i^{2}b_i^{2}}{a_i^{2}+b_i^{2}}\leq  \sum_{i=1}^{n}a_i^2 \sum_{i=1}^{n}b_i^2.
\end{equation}
where $\{a_i\}_{i=1}^n$, $\{b_i\}_{i=1}^n$ are two sequences of real numbers. Therefore
$$ (\Delta A)^2(\Delta B)^2= \langle \alpha|\alpha\rangle \langle \beta|\beta\rangle
=\sum_i^n |\alpha_i|^2\sum_i^n |\beta_i|^2\geq \sum_{i=1}^{n}(|\alpha_i|^{2}+|\beta_i|^{2})\sum_{i=1}^{n}\frac{|\alpha_i|^{2}|\beta_i|^{2}}{|\alpha_i|^{2}+|\beta_i|^{2}}.
$$

\textbf{Remark}
For observables $A$ and $B$, we can get the uncertainty relation in the following
\begin{equation}\label{th3}
(\Delta A)^2(\Delta B)^2\geq max\{\mathcal{L}_1, \mathcal{L}_2\}.
\end{equation}

Both the Callebaut and Milne inequalities (\ref{1}),(\ref{3})
are stronger than the usual Cauchy-Schwarz inequality, thus
our uncertainty relations are tighter than those uncertainty relations derived from the Cauchy-Schwaz inequality, for example, the
uncertainty inequality in \cite{pati} about observables:
\begin{equation}\label{pati1}
(\Delta A)^2(\Delta B)^2\geq\frac{1}{4}(\sum_n|\langle [\bar{A},\bar{B}_n^{\varphi}]\rangle _{\Psi}+\langle \{\bar{A},\bar{B}_n^{\varphi}\}\rangle _{\Psi}|)^2.
\end{equation}
where $\bar{B}_n^{\varphi}=|\varphi_n\rangle\langle\varphi_n|\bar{B}$. The following example shows the our bound
is tighter than the bound given by Mondal-Bagchi-Pati.

\textbf{Example:}We plot the lower bound of the product of variances of two incompatible observables, $A=L_x$, $B=L_y$, two components of the angular momentum of spin one particle with a state $|\Psi\rangle=cos\theta|1\rangle-sin\theta|0\rangle$, where the state $|1\rangle$ and $|0\rangle$ are the eigenvectors of $L_z$ corresponding to eigenvalues 1 and 0 respectively. Take the angular momentum operators with $\hbar=1$:
\begin{equation}
L_{x}=\frac{1}{\sqrt{2}}\left(
  \begin{array}{ccc}
    0 & 1 & 0 \\
    1 & 0 & 1\\
    0 & 1 & 0\\
  \end{array}
\right),
L_{y}=\frac{1}{\sqrt{2}}\left(
  \begin{array}{ccc}
    0 & -i & 0 \\
    i & 0 & -i\\
    0 & i & 0\\
  \end{array}
\right),
L_{z}=\left(
  \begin{array}{ccc}
    1 & 0 & 0 \\
    0 & 0 & 0\\
    0 & 0 & -1\\
  \end{array}
\right)
\end{equation}

\begin{figure*}[htbp]
\begin{center}
\includegraphics[scale=0.5]{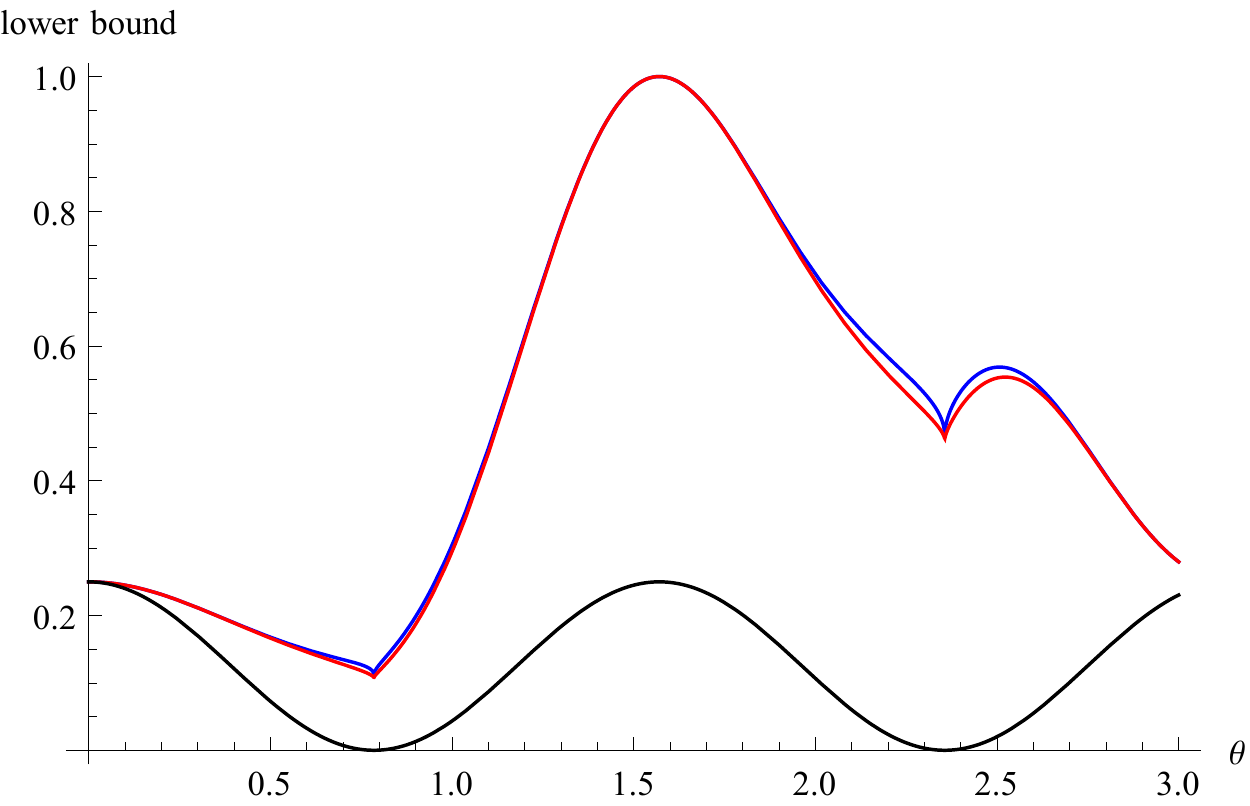}
\end{center}
\caption{The uncertainty relation for observables $L_{x}$, $L_{y}$ at state $|\Psi\rangle$: the blue curve is
the lower bound in (\ref{res1}) with the weight $\lambda=\frac{1}{2}$, the red one is with the weight $\lambda=\frac{1}{3}$, and the black one is for Mondal-Bagchi-Patis' lower bound in (\ref{pati1})\label{Fig:1}.}
\end{figure*}

\section{Conclusions}
Uncertainty relations play a central role in 
the current research in quantum theory and quantum information \cite{app1, app2, app3}.
We have derived a family of new product forms of variance-based uncertainty relations, which are
expected to help further investigate the uncertainty relation.

\bigskip
\noindent{\bf Acknowledgments}.
This work is supported by the National Natural Science Foundation of China under
Grant Nos. 11501153, 11461018 and 11531003; the Natural Science Foundation of Hainan Province
under Grant Nos. 20151010 and 20161006; the Scientific
Research Foundation for Colleges of Hainan Province under Grant
No. Hnky2015-18 and Simons Foundation grant 523868.

\end{document}